\RequirePackage[T1]{fontenc}
\documentclass[journal]{IEEEtran}
\usepackage{amsmath,amsfonts,amssymb}
\usepackage{algorithmic}
\usepackage{algorithm}
\usepackage{array}
\usepackage[caption=false,font=normalsize,labelfont=sf,textfont=sf]{subfig}
\usepackage{textcomp}
\usepackage{stfloats}
\usepackage{url}
\usepackage{graphicx}
\usepackage{cite}
\usepackage{booktabs}
\begin{document}

\title{Adversarial Pilot Injection Against AI-Based CSI Feedback in FDD Massive MIMO Systems}

\author{Jinli Zhai, Yusen Wang, Qimei Cui, Zengbao Zhu, Shengyuan Liang, and Xiaofeng Tao

\thanks{J. Zhai, Y. Wang, Q. Cui, Z. Zhu, S. Liang, and X. Tao are with the National Engineering Research Center for Mobile Network Technologies, Beijing University of Posts and Telecommunications, Beijing 100876, China. Q. Cui and X. Tao are also with the Department of Broadband Communication, Peng Cheng Laboratory, Shenzhen 518055, China. (e-mail: chenest@bupt.edu.cn; 2025110106@bupt.cn; cuiqimei@bupt.edu.cn; zengbaozhu@bupt.edu.cn; liangshengyuan@bupt.edu.cn; taoxf@bupt.edu.cn).}

\thanks{Corresponding author: Q. Cui (e-mail: cuiqimei@bupt.edu.cn).}}

\maketitle
\begin{abstract}
Artificial intelligence (AI)-based channel state information (CSI) feedback reduces overhead in frequency division duplexing (FDD) massive multiple-input multiple-output (MIMO) systems, yet its reconstructed CSI depends on the integrity of pilot acquisition. This paper identifies and evaluates the attack path of adversarial pilot injection against this feedback. A transmitted disturbance can damage the user-side estimation before encoding; correctly delivered quantized bits can still represent a contaminated channel estimate. We model this path through attacker propagation, linear minimum mean square error estimation, compression, quantization, and decoding under a total transmit-energy constraint. An adversarial perturbation generator (APG) implements the attack using channel statistics without current legitimate CSI or instantaneous attacker-channel observations at deployment. At a Rician attacker-channel factor of 10~dB and a transmit perturbation-to-signal ratio of $-4$~dB, APG produces a normalized mean-square error (NMSE) of 0.84~dB in base-station CSI reconstruction, compared with $-10.31$~dB for Gaussian interference. Estimator-targeted and reconstruction-targeted attacks yield different error orderings, and reconstruction damage also degrades beamforming. As a countermeasure, alternating adversarial training (AAT) adapts the feedback encoder and decoder to contaminated pilots. With condition-matched training and APGs reoptimized for the final receiver, AAT reduces NMSE by 2.40--3.79~dB across six representative channel--power conditions. The results demonstrate the importance of protecting pilot acquisition alongside quantized feedback delivery.
\end{abstract}
\begin{IEEEkeywords}
AI-based CSI feedback, massive MIMO, pilot injection, adversarial attack, adversarial training.
\end{IEEEkeywords}

\section{Introduction}
\subsection{Background}
Accurate downlink CSI enables the base station (BS) to align transmitted signals with the propagation channel and support coherent beamforming in massive MIMO systems. In FDD, the downlink and uplink occupy different frequency bands, so the BS obtains downlink CSI through pilot transmission and user equipment (UE) feedback rather than direct channel reciprocity. As the antenna array grows, the spatial and frequency dimensions of the channel increase the information that must be returned within a channel-coherence interval~\cite{guo2022overview}. The quality of feedback consequently affects both reconstruction accuracy and the usefulness of the resulting precoder.

AI-based CSI compression uses a UE encoder that maps channel information to a compact representation, and a BS decoder that reconstructs it~\cite{wen2018deep,cui2025overview}. Joint training exploits channel correlations to reduce feedback overhead while retaining information useful for transmission~\cite{guo2020convolutional}. The Third Generation Partnership Project (3GPP) includes CSI feedback enhancement, such as spatial - frequency compression using a two - sided AI model, in the technical report (TR) (38.843)~\cite{tr38843}. Quantization converts learned representations into finite-bit payloads. CSI quality needs to be maintained in acquisition, compression and reconstruction.

The reliability of the chain is not only related to the accuracy of neural networks under nominal observation. Adversarial perturbations can exploit the sensitivity of learned mappings~\cite{szegedy2013intriguing,chakraborty2021survey}, while wireless propagation changes the disturbance that reaches the receiver~\cite{wang2023adversarial,bahramali2021robust}. In CSI feedback, errors before compression will change the channel representation of downlink transmission channels. Evaluating the acquisition stage is necessary to connect learned feedback security with communication functions.

\subsection{Motivation}
The feedback link and the channel-acquisition stage expose different attack opportunities. Channel coding, cyclic redundancy check (CRC), and automatic repeat request (ARQ) can support correct delivery of a quantized payload. They do not determine whether the UE formed that payload from an accurate channel estimate. Attackers transmitting during downlink pilot intervals may contaminate the observations used by the estimator. The bits received at the BS may be correct, but they represent corrupted channel estimates. This separation opens up an attack path for artificial intelligence-based CSI feedback via pilot acquisition, even if the digital feedback link performs correctly.

Pilot contamination can manipulate channel training and redirect subsequent precoding~\cite{zhou2012pilot}. In the feedback of artificial intelligence, the attack effect relies on the mapping from the pilot observation to the CSI reconstruction after compression and quantization. This mapping can reduce or transform the estimation error, but the magnitude alone does not determine the final damage. We analyze transmitted pilot disturbances optimized by this complete mapping and evaluate their effects on channel direction and beamforming.

The information of the attacker is equally important. Optimizing known propagation can exploit the instantaneous structure, but it may be difficult to obtain this realization. Distribution optimization has another route: waveform generators can be trained over channels and deployed without observing legitimate CSI. Distribution-based optimization accounts for environmental variation in physical adversarial examples~\cite{athalye2018synthesizing}; here it incorporates attacker-to-UE propagation and acquisition. The key is whether the pilot injection path is effective with statistical channel knowledge, and whether its effect changes with instantaneous knowledge or statistical mismatches.

A defense must also address the attacker's response to the protected receiver. Adversarial training formulates this interaction through robust optimization~\cite{madry2018towards}, while adaptive-evaluation studies show why the attack must be tailored to the final defense~\cite{tramer2020adaptive}. For CSI feedback training, the training goal is to ensure that the reconstruction is carried out during normal operation. We explore whether training the feedback encoder and decoder on contaminated pilots can mitigate the identified threat and maintain the accuracy of normal operations. Attacks reoptimized against the final receiver assess this countermeasure, and controlled comparisons examine its reconstruction--beamforming tradeoff.

\subsection{Related Work}
\emph{Learned CSI acquisition and feedback:} Learned encoder--decoder representations exploit channel correlations to reduce feedback overhead~\cite{wen2018deep,li2020novel,lu2020multi,cui2022transnet,bi2022novel}, extending earlier codebook and compressive-sensing approaches~\cite{ziao2025review,kuo2012compressive}. Multiple-rate quantized feedback~\cite{guo2020convolutional} and joint estimation--compression or pilot-compression networks~\cite{guo2024joint} address finite bit budgets and imperfect observations. Security is related to how physically transmitted adversarial disturbances affect CSI after compression/quantized delivery. We use fixed estimators to examine this attack path, and optimize the disturbance before final reconstruction.

\emph{Attacks on CSI feedback and acquisition:} Liu et al. formulated encoder-input CSI perturbations and studied universal attacks on compressed analog feedback~\cite{liu2020adversarial}. The reconstruction loss reveals the sensitivity of the model, but does not elaborate on the transmission waveform that reaches the encoder via pilot acquisition. Zhu et al. found encoder-to-decoder link attacks ineffective under the considered conditions~\cite{zengbaoadversarial}. Other threats poison compressed beamforming reports~\cite{meneghello2024whack} or tamper with the reported transmission rank~\cite{timilsina2025tampering}. The injection precedes the encoding, so correctly delivered payloads can represent corrupted estimates.

Physical acquisition attacks influence precoding through pilot contamination~\cite{zhou2012pilot} or alter channel-quality and transmission-rank reports through selective interference~\cite{dino2026silent}. Hou et al. generated input-agnostic interference with estimated attacker-channel compensation~\cite{hou2022undermining}; Mendon\c{c}a et al. trained a channel estimator against regenerated pilot disturbances~\cite{mendonca2024jamming}. The unresolved issue here is how adversarial pilot injection affects CSI after AI-based compression and quantized delivery. We model this acquisition-to-feedback path and show that estimation-error and reconstruction-error rankings can differ under a common transmit-energy constraint. Statistical-channel evaluation further characterizes the path without instantaneous attacker-channel observations. The countermeasure fixes the estimator and trains the feedback encoder--decoder.

\emph{Input-agnostic wireless attacks:} Channel-aware attacks on classification and beam prediction account for wireless propagation~\cite{kim2021channel,sadeghi2018adversarial,kim2021adversarial}. Bahramali et al. incorporated communication constraints into perturbations requiring no target signal at deployment~\cite{bahramali2021robust}. SemAdv and SemMixed use random-input generation and mixed training for semantic communication~\cite{nan2023physical}. Here, training over known attacker-channel statistics supports generation without current legitimate CSI or instantaneous attacker-channel observations; paired knowledge and transfer comparisons examine its effects.

\emph{Robust training and adaptive evaluation:} The standard adversarial training retains the perturbations of the current model, and regenerates them~\cite{madry2018towards,bai2021recent}. Lee et al. alternated generator and classifier updates using clean and adversarial losses~\cite{lee2017generative}. We use this training family as a countermeasure for pilot injection into AI-based CSI feedback. Reoptimized attacks and differentiation through quantization follow adaptive-evaluation practice~\cite{athalye2018obfuscated,tramer2020adaptive}. The evaluation addresses whether the feedback network can recover useful CSI after contaminated acquisition, including the effect on beamforming.

\subsection{Contributions and Organization}
We investigate adversarial pilot injection into AI-based quantized CSI feedback, with APG and AAT implementing attack and mitigation, respectively. The contributions are:
\begin{enumerate}
\item We identify and formulate a physical attack path that contaminates downlink pilots before feedback encoding. The model links the transmitted disturbance through propagation, estimation, compression, quantization, and decoding to BS reconstruction under a total transmit-energy constraint. It exposes an acquisition-to-feedback threat despite correct payload delivery, whose damage depends on the complete reconstruction mapping.
\item We validate the attack path using reconstruction-targeted waveforms implemented by APG and a directly optimized universal grid. Estimation--reconstruction comparisons and beamforming metrics establish its communication consequences; paired channel-knowledge and fixed-source transfer tests characterize the conditions under which statistical-channel injection remains effective.
\item We evaluate feedback-network training as a countermeasure to the identified threat, implemented through AAT with the estimator fixed. Reoptimized attacks assess mitigation after attacker adaptation, while training controls and clean-loss comparisons establish its feasibility and reveal reconstruction--beamforming tradeoffs under the modeled statistical-channel threat.
\end{enumerate}
Section~\ref{sec:system} defines the acquisition and feedback model. Sections~\ref{sec:attack} and~\ref{sec:defense} describe the attack path and countermeasure, respectively. Section~\ref{sec:evaluation} presents the simulation settings and evidence for the three contributions, followed by the conclusion.

\section{System Model}
\label{sec:system}
\begin{figure*}[!t]
\centering
\includegraphics[width=\textwidth]{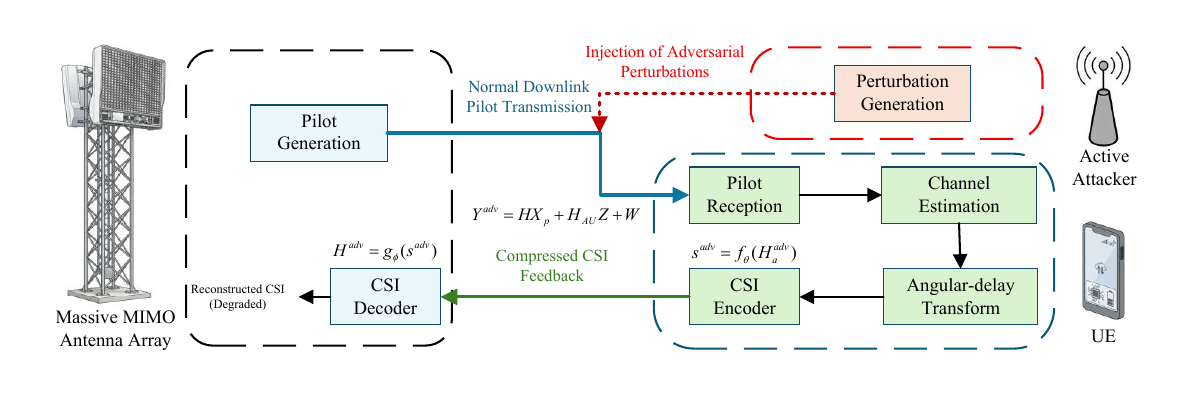}
\caption{Adversarial pilot injection into the CSI acquisition and feedback pipeline. The schematic illustrates contamination before feedback encoding; the pilot-observation and quantized-delivery model are specified in Section~\ref{sec:system}.}
\label{fig:system}
\end{figure*}

\subsection{Downlink Channel and Pilot Allocation}
We consider a BS with $N_t$ antennas and a single-antenna UE using orthogonal frequency-division multiplexing (OFDM) with $N_f$ subcarriers. The BS-to-UE channel is constant within a pilot block and has $N_c$ delay taps, with cyclic prefix (CP) length $N_{\rm CP}\geq N_c-1$. Its frequency response is
\begin{equation}
 H_{k,a}=\sum_{\ell=0}^{N_c-1}h_{\ell,a}
             e^{-j2\pi k\ell/N_f},
 \quad \mathbf H\in\mathbb C^{N_f\times N_t},
\label{eq:frequency}
\end{equation}
where $a=0,\ldots,N_t-1$ and $k=0,\ldots,N_f-1$.

One OFDM symbol carries frequency-interleaved pilots. Let $M=N_f/N_t$ and assign antenna $a$ the set
$\mathcal P_a=\{a+mN_t:m=0,\ldots,M-1\}$.
For a unit-modulus pilot sequence $q_c[m]$, selected from a known library using index $c$, the pilot grid is
\begin{equation}
 X_p[a,k;c]=
 \begin{cases}
 \sqrt{\rho_p}\,q_c[m],& k=a+mN_t,\\
 0,& k\notin\mathcal P_a .
 \end{cases}
\label{eq:pilots}
\end{equation}
Here $\rho_p>0$ is the power per active pilot tone. The grid $\mathbf X_p(c)\in\mathbb C^{N_t\times N_f}$ has disjoint antenna resources. This resource orthogonality does not require different library sequences to be mutually orthogonal. We require $M\geq N_c$ so that each antenna's pilot observations identify its modeled tap vector.

After CP removal and a unitary fast Fourier transform (FFT), the clean UE observation on subcarrier $k$ is
\begin{equation}
 y_k=\sum_{a=0}^{N_t-1}H_{k,a}X_p[a,k;c]+w_k,
 \qquad w_k\sim\mathcal{CN}(0,\sigma_w^2).
\label{eq:cleanobs}
\end{equation}
Collect the observations in antenna-resource order as
$\mathbf y=[\mathbf y_0^{T},\ldots,\mathbf y_{N_t-1}^{T}]^{T}$,
where $\mathbf y_a$ contains the samples on $\mathcal P_a$.
For $\mathbf h_a=[h_{0,a},\ldots,h_{N_c-1,a}]^T$, define
\begin{equation}
 [\mathbf B_a(c)]_{m,\ell}
  =\sqrt{\rho_p}\,q_c[m]\,
       e^{-j2\pi(a+mN_t)\ell/N_f}.
\label{eq:comboperator}
\end{equation}
Stacking $\mathbf h=[\mathbf h_0^T,\ldots,\mathbf h_{N_t-1}^T]^T$
and $\mathbf B_c=\operatorname{blkdiag}(\mathbf B_0,\ldots,\mathbf B_{N_t-1})$ gives
\begin{equation}
 \mathbf y=\mathbf B_c\mathbf h+\mathbf w.
\label{eq:stacked}
\end{equation}
Each $\mathbf B_a$ has full column rank for the stated tap support. Thus, the comb observes all modeled taps; the absolute subcarrier indices in~\eqref{eq:comboperator} also account for the antenna-dependent delay phase.

\subsection{LMMSE Channel Estimation}
The UE uses a tap-domain linear minimum mean square error (LMMSE) estimator with prior mean
$\boldsymbol\mu_h$ and covariance $\mathbf R_h$ estimated from legitimate training channels. With
\begin{equation}
 \mathbf K_c=\mathbf R_h\mathbf B_c^H
   \big(\mathbf B_c\mathbf R_h\mathbf B_c^H+
                       \sigma_w^2\mathbf I\big)^{-1},
\label{eq:lmmsegain}
\end{equation}
the estimate is
\begin{equation}
 \widehat{\mathbf h}_{\rm UE}
  =\boldsymbol\mu_h+\mathbf K_c
       (\mathbf y-\mathbf B_c\boldsymbol\mu_h).
\label{eq:lmmse}
\end{equation}
The corresponding $\widehat{\mathbf H}_{\rm UE}$ follows from~\eqref{eq:frequency}. The prior and noise variance are fixed before attack and defense training. They are not recomputed from attacked observations or test labels. The pilot signal-to-noise ratio (SNR) is defined using the average clean received pilot energy:
\begin{equation}
 {\rm SNR}_p=
 10\log_{10}
 \frac{\mathbb E_{\rm train,c}
          \|\mathbf B_c\mathbf h\|_2^2}
      {N_f\sigma_w^2}.
\label{eq:snr}
\end{equation}
The noise variance is fixed using this training-set calibration. Injection changes the received interference level, rather than the reference SNR.

\subsection{Quantized CSI Feedback}
Let $\mathbf F_N$ denote the unitary discrete Fourier transform (DFT) matrix with
$[\mathbf F_N]_{r,s}=N^{-1/2}e^{-j2\pi rs/N}$, and let
$\mathbf S\in\mathbb R^{N_c\times N_f}$ select the first $N_c$ delay bins. The true and estimated angular-delay representations are
\begin{align}
 \mathbf A&=\mathbf S\mathbf F_{N_f}^{H}
                 \mathbf H\mathbf F_{N_t},\label{eq:angulartrue}\\
 \widetilde{\mathbf A}_{\rm UE}
   &=\mathbf S\mathbf F_{N_f}^{H}
       \widehat{\mathbf H}_{\rm UE}\mathbf F_{N_t}.
\label{eq:angularue}
\end{align}
Real and imaginary components of $\widetilde{\mathbf A}_{\rm UE}$
form the encoder input, scaled by a single training-set root-mean-square (RMS) factor
$\gamma^2=\mathbb E_{\rm train}\|\mathbf A\|_F^2/(N_cN_t)$.
This scale is shared across data splits and receiver variants.

A Transformer encoder~\cite{vaswani2017attention} produces $d_s$ real latent values:
\begin{equation}
 \mathbf u=f_e(\widetilde{\mathbf A}_{\rm UE}/\gamma;\theta_e),
 \qquad \mathbf u\in[-1/2,1/2]^{d_s}.
\end{equation}
The complex-input notation here includes real/imaginary concatenation. Each scalar is uniformly quantized with $b$ bits:
\begin{align}
 Q_b(u)&=\operatorname{clip}\big(\lfloor 2^b(u+1/2)\rfloor,0,2^b-1\big),\\
 U_b(j)&=(j+1/2)/2^b-1/2.
\end{align}
The index $j$ is represented by $b$ bits, and $U_b$ maps it to its quantization-bin center. The CSI payload has $B_{\rm fb}=bd_s$ bits, excluding CRC, coding, and retransmission overhead.

We assume correct delivery of these bits, supported by CRC and ARQ, within the considered channel coherence interval. Quantization is part of this model; feedback-bit errors are excluded. The BS decoder reconstructs
\begin{equation}
 \widehat{\mathbf A}
  =\gamma f_d\big(U_b(Q_b(\mathbf u));\theta_d\big).
\end{equation}
Zero-padding the delay dimension gives
\begin{equation}
 \widehat{\mathbf H}_{\rm BS}
   =\mathbf F_{N_f}\mathbf S^T
                \widehat{\mathbf A}\mathbf F_{N_t}^{H}.
\label{eq:inverse}
\end{equation}
The ground truth $\mathbf H$, UE estimate $\widehat{\mathbf H}_{\rm UE}$,
and BS reconstruction $\widehat{\mathbf H}_{\rm BS}$ are distinct. For the modeled tap support,~\eqref{eq:angulartrue} and~\eqref{eq:inverse} are inverse transforms without truncation loss.

Write the complete receiver and feedback mapping as
$\widehat{\mathbf H}_{\rm BS}=F_{\theta_f}(\mathbf y;c)$,
where $\theta_f=(\theta_e,\theta_d)$ are the trainable parameters.
Clean pretraining and reconstruction-targeted attack optimization use
\begin{equation}
 \ell(\mathbf H,\widehat{\mathbf H})
   =\frac{\|\mathbf H-\widehat{\mathbf H}\|_F^2}
          {\max\{\|\mathbf H\|_F^2,\epsilon\}},
\label{eq:loss}
\end{equation}
with a fixed numerical floor $\epsilon>0$, as in the implementation. The reconstruction target is the true legitimate CSI, rather than the clean estimator output.

\section{Adversarial Pilot Injection Against CSI Feedback}
\label{sec:attack}
\subsection{Attack Surface and Knowledge}
The attacker has $N_a$ transmit antennas and injects an OFDM-compatible signal during the downlink pilot symbol. Its CP covers the modeled attacker-channel delay spread. Both links are block-fading, with timing and carrier synchronization during this symbol. Let
$\mathbf Z\in\mathbb C^{N_a\times N_f}$ be the frequency-domain attack grid and $\mathbf G\in\mathbb C^{N_f\times N_a}$ the attacker-to-UE frequency response. The attacked observation is
\begin{equation}
 y_k^{\rm adv}=
 \sum_a H_{k,a}X_p[a,k;c]
 +\sum_b G_{k,b}Z[b,k]+w_k.
\label{eq:advobs}
\end{equation}
Define $\mathcal J_{\mathbf G}(\mathbf Z)$ as the vector of interference samples $\sum_bG_{k,b}Z[b,k]$, ordered as in~\eqref{eq:stacked}. Then
\begin{equation}
 \mathbf y^{\rm adv}=\mathbf B_c\mathbf h+
           \mathcal J_{\mathbf G}(\mathbf Z)+\mathbf w.
\label{eq:advstack}
\end{equation}
The fixed estimator satisfies
\begin{equation}
 \widehat{\mathbf h}_{\rm UE}^{\rm adv}
 -\widehat{\mathbf h}_{\rm UE}
   =\mathbf K_c\mathcal J_{\mathbf G}(\mathbf Z)
\label{eq:estimatorbias}
\end{equation}
for the same legitimate channel and noise realization. This links the transmitted signal to the feedback-network input.

We consider white-box access to the deployed encoder and decoder, pilot schedule and sequence, LMMSE parameters, quantizer, and preprocessing. The attacker can use independent legitimate training channels offline. At deployment, the statistical-channel APG uses a random input and known pilot information for power projection; it does not observe the current $\mathbf H$, $\mathbf h$, noise realization, or clean received pilots. The channel-known benchmark additionally knows the current $\mathbf G$. No attack is injected into the digital feedback link.

\subsection{Transmit-Energy Constraint}
Let $\mathcal T_{\rm CP}$ apply a unitary inverse FFT to every transmit antenna and append its CP. The legitimate energy over the entire pilot transmission window is
\begin{equation}
 E_p(c)=\|\mathcal T_{\rm CP}(\mathbf X_p(c))\|_F^2.
\end{equation}
For a perturbation-to-signal ratio (PSR) budget $\eta$ in dB, the feasible attack set is
\begin{equation}
 \begin{aligned}
 \mathcal Z_{\eta,c}
   &=\left\{\mathbf Z:
       \|\mathcal T_{\rm CP}(\mathbf Z)\|_F^2
       \leq E_a(c)\right\},\\
 E_a(c)&=10^{\eta/10}E_p(c).
 \end{aligned}
\label{eq:budget}
\end{equation}
All antennas have the same CP interval, and PSR is the total transmit energy ratio. The received interference power relies on $\mathbf G$ to be recorded as an outcome. The waveform does not renormalize using instantaneous receiving power.

The feasible output of a raw grid $\mathbf V$ is
\begin{equation}
 \Pi_{\eta,c}(\mathbf V)=
  \mathbf V\min\left\{1,
   \frac{\sqrt{E_a(c)}}
        {\|\mathcal T_{\rm CP}(\mathbf V)\|_F}\right\},
\label{eq:projection}
\end{equation}
with $\Pi_{\eta,c}(\mathbf 0)=\mathbf 0$. Radial rescaling ensures the energy constraint in each forward pass.

\subsection{Attacker Channel Statistics}
The attacker channel has $N_g$ delay taps with $N_g-1\leq N_{\rm CP}$. Its impulse response at tap $\ell$ is
\begin{equation}
 \mathbf g_\ell=\sqrt{\beta_A}
 \left(\sqrt{\frac{\kappa}{\kappa+1}}\,\mathbf v_\ell+
       \sqrt{\frac{1}{\kappa+1}}\,\mathbf s_\ell\right),
\label{eq:rician}
\end{equation}
where $\mathbf g_\ell,\mathbf v_\ell,\mathbf s_\ell\in\mathbb C^{N_a}$.
The line-of-sight (LOS) component is at tap zero:
$[\mathbf v_0]_b=e^{j(\phi_0+\pi b\sin\vartheta_0)}$ for a half-wavelength uniform linear array, and $\mathbf v_\ell=\mathbf 0$ for $\ell>0$.
The scattered taps are independent with
$\mathbf s_\ell\sim\mathcal{CN}(\mathbf 0,p_\ell\mathbf I)$ and
\begin{equation}
 p_\ell=\frac{e^{-\ell/\tau_A}}
                 {\sum_{r=0}^{N_g-1}e^{-r/\tau_A}},
 \qquad \ell=0,\ldots,N_g-1.
\label{eq:pdp}
\end{equation}
Here $\tau_A>0$ controls the scattered delay-power decay. The frequency response follows~\eqref{eq:frequency} with the attacker tap support $N_g$ in place of $N_c$.

For $\kappa>0$, $K_{\rm dB}=10\log_{10}\kappa$.
Pure non-line-of-sight (NLOS) propagation is modeled by $\kappa=0$; $K_{\rm dB}=0$ instead means equal LOS and scattered power. The average gain is constant across $K$:
$\mathbb E\sum_\ell\|\mathbf g_\ell\|_2^2=N_a\beta_A$.
Individual channel realizations retain their sampled gains. The mean legitimate-link gain is $\beta_B=\mathbb E_{\rm train}\|\mathbf h\|_2^2/N_t$; Section~\ref{subsec:simulation_setup} specifies the relative link gain and LOS parameters used in the experiments.

Denote the full distribution by $p_\xi(\mathbf G)$.
The statistical knowledge $\xi$ contains the average gain, tap-power profile, LOS direction and phase, $K$-factor, and spatial covariance. Changing $K$ alone therefore does not represent uncertainty in every channel parameter.

\subsection{Attack Implementation With APG}
We use APG to implement a reconstruction-targeted pilot-injection attack. A one-dimensional convolutional generator maps
$\mathbf n\sim\mathcal N(\mathbf 0,\mathbf I)$ to a raw grid:
\begin{equation}
 \mathbf Z_{\theta_g}(\mathbf n;c)
       =\Pi_{\eta,c}\big(C_{\psi}(\mathbf n)+\mathbf T\big).
\label{eq:generator}
\end{equation}
Here $C_{\psi}$ is a convolutional mapping with normalization and nonlinear activation, and $\mathbf T\in\mathbb C^{N_a\times N_f}$ is a trainable frequency-specific template, initialized to zero. Together they form $\theta_g=(\psi,\mathbf T)$. The convolutional mapping and template jointly parameterize the transmitted grid. Real and imaginary output channels form the complex grid; $\mathcal T_{\rm CP}$ converts it into the transmitted waveform. The random input is resampled independently for each generated waveform. Neither component receives the current clean pilot observation or test CSI. Normalization statistics are updated only during generator training and remain fixed during validation and deployment.

With the feedback network fixed, define
\begin{equation}
 \begin{split}
 \mathcal R_q(\theta_f,\theta_g)
  =\mathbb E_{\mathbf H,c,\mathbf w,\mathbf n,\mathbf G\sim q}
   \big[\ell\big(\mathbf H,\,
    F_{\theta_f}(\mathbf B_c\mathbf h\\
       +\mathcal J_{\mathbf G}
           (\mathbf Z_{\theta_g}(\mathbf n;c))
       +\mathbf w;c)\big)\big].
 \end{split}
\label{eq:attackrisk}
\end{equation}
Here $\mathbf H$ comes from the legitimate training distribution, $c$ is sampled from the pilot library, and channel and noise samples are independent of $\mathbf n$. APG training approximately maximizes $\mathcal R_q$ through stochastic gradient ascent, with~\eqref{eq:projection} enforcing feasibility.

For the \emph{channel-known attack}, $q$ is a point mass at a known realization $\mathbf G^0$. The APG is optimized and tested on that realization, while legitimate test CSI remains held out. This is an ideal optimization benchmark for a coherent attacker channel. To isolate the value of channel knowledge, separate channel-known APGs are optimized on a common set of attacker-channel realizations and compared with a statistical APG on those same realizations.

For the \emph{statistical-channel attack}, $q=p_\xi$. Fresh attacker channels are drawn during optimization, following a distribution-based attack objective~\cite{athalye2018synthesizing}, and test channels are independent draws from the specified distribution. One APG is used across those test realizations. Channel transfer instead fixes the APG trained under $\xi_{\rm src}$ and evaluates it under $\xi_{\rm test}$ without retraining.

A directly optimized universal adversarial perturbation (UAP) provides a second waveform parameterization for evaluating the same attack path under a common information and energy budget.

\subsection{Pilot-Shaped Contamination}
Following the pilot-contamination principle~\cite{zhou2012pilot}, a physical baseline is $\mathbf Z=\alpha\mathbf X_p(c)$ with
$|\alpha|^2=10^{\eta/10}$ and $N_a=N_t$. This satisfies~\eqref{eq:budget} and gives the effective pilot channel
\begin{equation}
 H_{k,a}^{\rm eff}=H_{k,a}+\alpha G_{k,a}.
\label{eq:pilotcontamination}
\end{equation}
The pilot observations identify this combined response. More generally, \eqref{eq:estimatorbias} maps injection into estimator error, whereas the nonlinear encoder, quantizer, and decoder determine its final reconstruction effect. Maximizing estimator error and maximizing reconstruction error therefore define different attack objectives. Statistical reconstruction priors can reduce expected error under the modeled distribution, while separation of arbitrary legitimate and contaminating channels requires additional information. Accordingly, AAT targets lower reconstruction risk under specified attack distributions.

\section{Countermeasure Through Feedback Training}
\label{sec:defense}
We use generator-based adversarial training~\cite{lee2017generative} as a countermeasure to pilot injection. AAT implements this countermeasure by retaining an APG throughout feedback-network training and updating both models through the acquisition--feedback mapping. The estimator remains fixed, while the encoder and decoder learn to reconstruct the legitimate CSI from clean and contaminated observations.

\subsection{Training Objective for Contaminated Pilots}
Let the clean reconstruction risk be
\begin{equation}
 \mathcal R_{\rm cl}(\theta_f)
 =\mathbb E_{\mathbf H,c,\mathbf w}
  \big[\ell(\mathbf H,
   F_{\theta_f}(\mathbf B_c\mathbf h+\mathbf w;c))\big].
\end{equation}
Following the robust-optimization formulation of adversarial training~\cite{madry2018towards}, the generator-based reconstruction objective is
\begin{equation}
 \min_{\theta_f}\max_{\theta_g}\;
 \lambda\mathcal R_{\rm cl}(\theta_f)
 +(1-\lambda)\mathcal R_{p_\xi}(\theta_f,\theta_g),
 \quad 0\leq\lambda<1,
\label{eq:aat}
\end{equation}
where all generated grids belong to $\mathcal Z_{\eta,c}$ through~\eqref{eq:generator}. The maximization acts on a shared generator over $p_\xi$, consistent with the statistical deployment threat in Section~\ref{sec:attack}. The weight $\lambda$ controls the emphasis on normal-operation accuracy; its effect is examined in Section~\ref{subsec:ablation}. The LMMSE prior, estimator, pilot allocation, transforms, and normalization remain fixed.

Hard quantization and dequantization are applied in every forward pass, including evaluation. During gradient-based optimization, a straight-through estimator (STE)~\cite{bengio2013estimating} approximates the derivative of the quantized latent mapping. The loss is computed against the true legitimate channel. This retains the acquisition and quantization effects within the adversarial objective.

\subsection{Generator--Receiver Updates}
Starting from a clean pretrained feedback network, each mini-batch alternates $n_g$ generator updates and one feedback-network update. Write the sampled adversarial risk as $\widehat{\mathcal R}_{\rm adv}$. A generator ascent step is
\begin{equation}
 \theta_g\leftarrow\theta_g+
       \alpha_g\nabla_{\theta_g}
        \widehat{\mathcal R}_{\rm adv}(\theta_f,\theta_g).
\end{equation}
The feedback update is
\begin{equation}
 \theta_f\leftarrow\theta_f-
       \alpha_f\nabla_{\theta_f}
       \big[\lambda\widehat{\mathcal R}_{\rm cl}
        +(1-\lambda)\widehat{\mathcal R}_{\rm adv}\big].
\end{equation}
Here $\alpha_g$ and $\alpha_f$ denote the generator and feedback-network learning rates, respectively. The alternating optimizer steps implement a stochastic approximation to~\eqref{eq:aat}.

When the generator updates, the feedback network parameters are frozen, and the gradients pass through the receiver to reach the transmitted grid. When reconstructing the observations of each update, fresh random inputs, channels and noise are used. A frozen network uses evaluation mode so that its running statistics do not change during the opposing update. Algorithm~\ref{alg:aat} summarizes the procedure.

\begin{algorithm}[t]
\caption{Feedback Training Against Pilot Injection}
\label{alg:aat}
\begin{algorithmic}[1]
\REQUIRE Training CSI, pilot library, fixed LMMSE statistics,
$p_\xi$, PSR $\eta$, clean weight $\lambda$, ascent count $n_g$
\STATE Initialize $\theta_f$ from clean pretraining and initialize $\theta_g$
\FOR{each training mini-batch $\mathbf H$}
 \FOR{$r=1,\ldots,n_g$}
  \STATE Freeze $\theta_f$; sample $c,\mathbf G,\mathbf w,\mathbf n$
  \STATE Generate feasible $\mathbf Z_{\theta_g}$ by~\eqref{eq:generator}
  \STATE Form $\mathbf y^{\rm adv}$ by~\eqref{eq:advstack}
  \STATE Update $\theta_g$ by ascent on the true-CSI loss
 \ENDFOR
 \STATE Freeze $\theta_g$; resample $c,\mathbf G,\mathbf w,\mathbf n$
 \STATE Generate and detach a feasible attack grid
 \STATE Form clean and attacked observations
 \STATE Update $\theta_f$ on the mixed loss in~\eqref{eq:aat}
\ENDFOR
\RETURN Encoder--decoder parameters $\theta_f$
\end{algorithmic}
\end{algorithm}

\subsection{Evaluation Against Adaptive Attacks}
The frozen attack shared by clean and defended networks is used to measure the same corruption response. There is a separate adaptive evaluation that freezes the final feedback network, and reoptimizes the fresh attacker. This separation follows adaptive-evaluation practice~\cite{tramer2020adaptive}: an APG trained against the original network need not remain the strongest attack on its defended counterpart.

The same transmit energy and channel information conditions are used for optimized APG and universal waveform attacks. Evaluation retains the actual hard-quantization forward path, with the STE used only during attack optimization~\cite{athalye2018obfuscated}.

\section{Simulation Results and Analysis}
\label{sec:evaluation}

This part evaluates the pilot injection path through reconstruction damage, beam-forming consequences, and channel information dependence. APG and universal-grid optimization implement reconstruction-targeted attacks, while AAT implements a feedback-training countermeasure. The analysis then assesses mitigation against adaptive attacks and the associated reconstruction--beamforming tradeoffs.

\subsection{Simulation Setup}
\label{subsec:simulation_setup}

We consider a downlink system with a 32-antenna BS and a single-antenna UE. The legitimate link follows the clustered delay line model C (CDL-C) in 3GPP TR~38.901~\cite{tr38901} with a nominal delay spread of 100~ns and remains static within each transmission block. The BS uses a half-wavelength linear array with isotropic, vertically polarized elements. Fixed profile-power scaling is used without per-realization channel normalization, and finite-sinc sampling produces the 27-tap discrete support. The training, validation, and test sets contain 32\,768, 4096, and 8192 independent channel realizations, respectively. The 32-antenna attacker communicates with the UE through the multipath model in~\eqref{eq:rician}. K10 and K4 denote attacker-channel Rician factors of 10 and 4~dB, respectively; pure NLOS contains no direct component. The legitimate-channel distribution remains fixed throughout these comparisons, so changes in performance reflect the attack and its propagation conditions.

\begin{figure*}[!t]
\centering
\includegraphics[width=\textwidth]{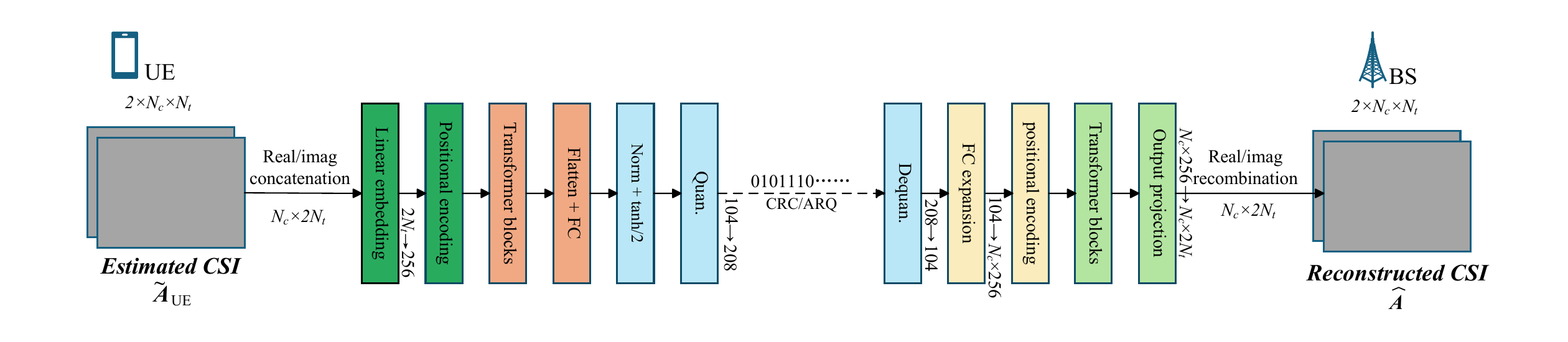}
\caption{Transformer CSI feedback architecture used in the experiments. The LMMSE angular-delay estimate forms $N_c$ tokens with $2N_t$ real features. Fully connected (FC) mappings compress and expand the latent representation. Layer Norm, Quan., and Dequan. denote layer normalization, quantization, and dequantization, respectively. The channel between them delivers the quantized payload through CRC/ARQ.}
\label{fig:architecture}
\end{figure*}

The UE performs LMMSE estimation from frequency-interleaved pilots and encodes the estimate using the quantized Transformer feedback network. Estimator priors and normalization parameters are determined from the training set and remain fixed. The feedback payload is delivered correctly through CRC/ARQ within the channel-coherence interval. All evaluated perturbations enter through the downlink pilot observations; the resulting payload is formed from the contaminated estimate. PSR follows~\eqref{eq:budget}: it is the ratio of total attack energy to legitimate pilot energy over the same transmission window, including the CP and all transmit antennas. Received interference depends on the waveform and attacker channel, so equal transmit PSR does not imply equal interference energy at the UE. This common transmit constraint allows the comparison to retain the effect of propagation and waveform design. Table~\ref{tab:parameters} summarizes the principal settings, and Fig.~\ref{fig:architecture} shows the feedback architecture.

\begin{table}[!t]
\caption{Principal Simulation Parameters}
\label{tab:parameters}
\centering
\small
\setlength{\tabcolsep}{3pt}
\begin{tabular}{@{}p{.56\columnwidth}p{.39\columnwidth}@{}}
\toprule
Parameter & Value\\
\midrule
BS / attacker antennas & 32 / 32\\
Legitimate-channel model & 3GPP CDL-C\\
Carrier frequency / subcarrier spacing & 2~GHz / 15~kHz\\
Subcarriers / CP length & 1024 / 32\\
Legitimate / attacker delay taps & 27 / 13\\
Pilot tones per BS antenna & 32\\
Clean pilot SNR & 10~dB\\
Latent dimension / quantization & 104 / 2~bits per scalar\\
Feedback payload & 208~bits\\
Transformer blocks / width / heads & 4 / 256 / 8\\
Feed-forward width / dropout & 512 / 0.1\\
Pretraining optimizer / learning rate & AdamW / $10^{-3}$\\
AAT feedback learning rate & $5\times10^{-5}$\\
Training batch size & 256\\
Pretraining / attack epochs & 288 / 30\\
APG convolution widths & 2 / 64 / 128 / 128 / 64\\
APG convolution kernels & 15 / 9 / 7 / 5\\
APG mapping / template learning rates & $2\times10^{-4}$ / $2\times10^{-2}$\\
APG steps per feedback step & 2\\
\bottomrule
\end{tabular}
\end{table}

The pilots use quadrature phase-shift keying (QPSK) symbols. The attacker delay profile has $\tau_A=3$ taps, with LOS direction and phase $\vartheta_0=\phi_0=0$ and equal average link gains $\beta_A=\beta_B$. APG combines a convolutional mapping with the learned frequency template in~\eqref{eq:generator}; its first three convolutions use batch normalization and leaky rectified linear activations. Encoder and decoder each contain the Transformer blocks in Table~\ref{tab:parameters}. Clean pretraining uses AdamW with a cosine learning-rate schedule and weight decay 0.01. APG and AAT feedback optimization use Adam; AAT uses $\lambda=0.5$ unless specified otherwise.

The attack baselines are Gaussian interference, pilot-shaped contamination, statistical beam jamming, LMMSE-targeted UAP (LMMSE-UAP), and reconstruction-targeted UAP optimized by projected gradient descent (PGD), denoted UAP-PGD~\cite{madry2018towards,bahramali2021robust}. Gaussian interference is spatially unstructured, pilot-shaped contamination reuses the legitimate pilot grid, and statistical beam jamming sends Gaussian frequency-domain symbols along the conjugate steering direction of the known LOS component. All satisfy the same CP-inclusive total transmit-energy constraint. LMMSE-UAP maximizes the energy of the attack-induced estimate displacement in~\eqref{eq:estimatorbias}, normalized by the true tap-vector energy, without querying the feedback network. UAP-PGD and APG target final CSI reconstruction. The reported UE NMSE evaluates the complete estimate against true CSI. Statistical APG uses known attacker-channel statistics and requires no current legitimate CSI, instantaneous attacker channel, or received pilots at deployment.

APG and UAP-PGD each use 30 training epochs with batch size 256. UAP-PGD takes normalized-gradient ascent steps of length $0.2\sqrt{E_a}$ followed by projection onto the transmit-energy constraint. The step size is selected on validation data. APG and UAP-PGD are selected by their largest validation reconstruction loss; LMMSE-UAP is selected by its largest validation estimator-displacement loss. Test data are used only for evaluation. Adaptive attacks are independently reoptimized against each final receiver under the same information and energy conditions.

For the defense comparison, a separate AAT receiver is trained for each evaluated attacker-channel distribution and PSR. Clean denotes the clean-pretrained receiver in Fig.~\ref{fig:defenseresults} and Table~\ref{tab:defense}; AAT further adapts that receiver to contaminated pilots. Defense training uses validation-plateau stopping. The ablation in Fig.~\ref{fig:ablation} instead continues every strategy, including Clean, from a common pretrained receiver to a common stopping epoch within each condition. It matches feedback-network training and validation-selection opportunities; alternating strategies additionally optimize APG. The final-system comparison measures achieved performance, while the ablation examines training effects under matched feedback training. A shared APG is optimized against Clean; adaptive evaluation reoptimizes the attacker for the final AAT receiver.

Reconstruction accuracy is measured by NMSE, directional fidelity by the squared generalized cosine similarity (SGCS), and beamforming performance by the maximum-ratio transmission (MRT) rate. For the true channel $\mathbf H_i$ and its reconstruction $\widehat{\mathbf H}_i$,
\begin{equation}
 \mathrm{NMSE}_{\mathrm{dB}}=10\log_{10}\!\left[
 \frac{1}{N}\sum_{i=1}^{N}
 \frac{\|\mathbf H_i-\widehat{\mathbf H}_i\|_F^2}
      {\|\mathbf H_i\|_F^2}\right].
\label{eq:nmse}
\end{equation}
Let $\mathbf h_{i,k}$ and $\widehat{\mathbf h}_{i,k}$ be the true and reconstructed spatial channel vectors on subcarrier $k$. Then
\begin{equation}
 \mathrm{SGCS}=\frac{1}{NN_f}\sum_{i,k}
 \frac{|\mathbf h_{i,k}^{H}\widehat{\mathbf h}_{i,k}|^2}
 {\|\mathbf h_{i,k}\|_2^2\|\widehat{\mathbf h}_{i,k}\|_2^2},
\label{eq:sgcs}
\end{equation}
and
\begin{equation}
 R_{\mathrm{MRT}}=\frac{1}{NN_f}\sum_{i,k}
 \log_2\!\left(1+\rho_d
 \frac{|\mathbf h_{i,k}^{H}\widehat{\mathbf h}_{i,k}|^2}
      {\|\widehat{\mathbf h}_{i,k}\|_2^2}\right).
\label{eq:rate}
\end{equation}
The common $\rho_d$ is calibrated from training channels so that perfect-CSI MRT yields an average received SNR of 10~dB. A zero reconstruction contributes zero to SGCS and rate. The rate measures the beamforming consequence of reconstructed CSI and excludes feedback and protocol overhead. NMSE measures the complete reconstruction discrepancy, whereas SGCS is invariant to a nonzero scalar rescaling of the reconstructed vector. Together, SGCS and MRT determine whether a reconstruction improvement preserves the spatial direction needed for transmission.

Continuous curves show complete parameter sweeps from one independent training trial; tables, grouped bars, and the stated repeated-trial comparisons summarize three independently initialized trials. Metrics are averaged within each trial and then equally across trials, with NMSE averaged in the linear domain before conversion to dB. The $\pm$ denotes the sample standard deviation of the trial-level linear means; error bars show the observed range across the trials. All methods use common test channels and observation conditions.

\subsection{Communication Impact of Pilot Injection}
\label{subsec:attack_effect}

\begin{figure*}[!t]
\centering
\includegraphics[width=\textwidth]{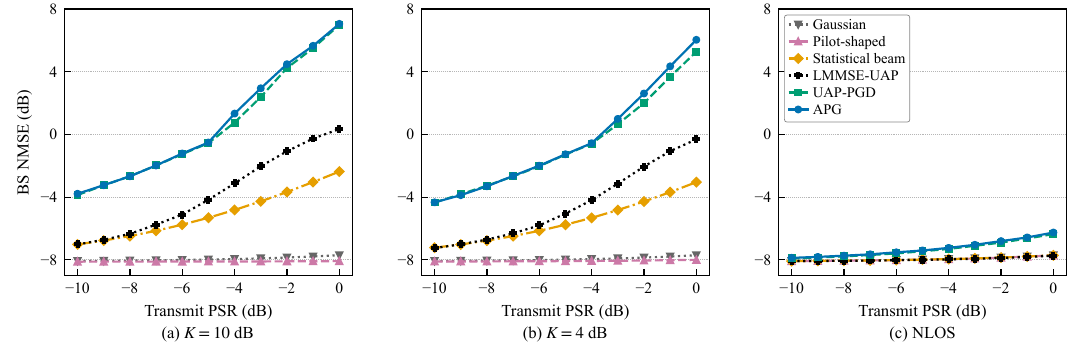}
\caption{BS reconstruction NMSE versus transmit PSR under (a) K10, (b) K4, and (c) NLOS attacker channels. All attack methods satisfy the same total transmit-energy constraint. Sweeps use one independent training trial; Table~\ref{tab:attack} summarizes three trials.}
\label{fig:attackresults}
\end{figure*}

Pilot injection optimized through the feedback mapping causes substantial reconstruction damage despite correct payload delivery. Under the same transmit-energy constraint, this damage exceeds that from unstructured interference. Fig.~\ref{fig:attackresults} shows the PSR sweeps in K10, K4, and NLOS. APG and UAP-PGD become increasingly destructive with PSR in K10 and K4, while Gaussian and pilot-shaped interference produce much smaller errors. Table~\ref{tab:attack} supplies the representative comparison at K10 and PSR $-4$~dB. Without injection, the same pretrained receiver gives a BS NMSE of $-10.83$~dB. APG raises it to 0.84~dB, compared with $-10.31$~dB for Gaussian interference and $-4.45$~dB for statistical beam jamming.

The additional separation from statistical beam jamming shows the value of selecting the waveform through the reconstruction objective as well as exploiting propagation structure. At this condition, the received attack energies of APG and statistical beam jamming, averaged across the repeated trials, differ by approximately 0.10~dB, whereas their BS NMSEs differ by 5.30~dB. Their similar received energies further support the role of waveform structure in the end-to-end reconstruction damage.

\begin{table}[!t]
\caption{Estimation and Reconstruction Performance at K10: No Attack and Injection With PSR $-4$~dB}
\label{tab:attack}
\centering
\footnotesize
\setlength{\tabcolsep}{3pt}
\begin{tabular}{@{}lrrrr@{}}
\toprule
Method & UE NMSE & BS NMSE & SGCS & MRT rate\\
 & (dB) & (dB) & & (bit/s/Hz)\\
\midrule
No attack & -22.75 & -10.83 & 0.921 & 3.290\\
Gaussian & -16.30 & -10.31 & 0.911 & 3.277\\
Pilot-shaped & -21.52 & -10.79 & 0.920 & 3.289\\
Statistical beam jamming & -2.73 & -4.45 & 0.728 & 2.988\\
LMMSE-UAP & 5.63 & -1.82 & 0.570 & 2.662\\
UAP-PGD & 2.89 & 0.62 & 0.480 & 2.421\\
APG & 2.80 & 0.84 & 0.485 & 2.429\\
\bottomrule
\end{tabular}\space\null
\end{table}

The estimation and reconstruction errors do not follow the same ordering even for attacks with a common waveform parameterization. LMMSE-UAP and UAP-PGD both optimize a universal transmit grid under the same information, energy, and training-budget conditions. LMMSE-UAP yields a UE NMSE of 5.63~dB, exceeding the 2.89~dB produced by UAP-PGD, but its BS NMSE is $-1.82$~dB, compared with 0.62~dB for UAP-PGD. The former optimizes the estimator displacement, whereas the latter accounts for the encoder, quantizer, and decoder in its reconstruction objective. Under the common waveform class, the final reconstruction is an important part of the pilot attack objective.

APG and UAP-PGD attain similar BS errors at this representative condition, with NMSE values of 0.84 and 0.62~dB, respectively. These two implementations confirm that the pilot-injection path can disrupt reconstruction when the waveform is optimized through the complete feedback mapping. The following comparisons use APG to examine the channel information and statistical matching on which the attack effect depends.

The reconstruction damage also affects beamforming. In Table~\ref{tab:attack}, the no-attack reference gives an SGCS of 0.921 and an MRT rate of 3.290~bit/s/Hz. Gaussian interference gives 0.911 and 3.277~bit/s/Hz, whereas APG reduces them to 0.485 and 2.429~bit/s/Hz. Because SGCS removes scalar gain differences, its reduction identifies a loss of channel-direction information. The associated MRT decrease shows the consequence for transmission using the reconstructed channel. Thus, the attack effect extends beyond an increase in the reconstruction-error metric.

\begin{figure*}[!t]
\centering
\includegraphics[width=\textwidth]{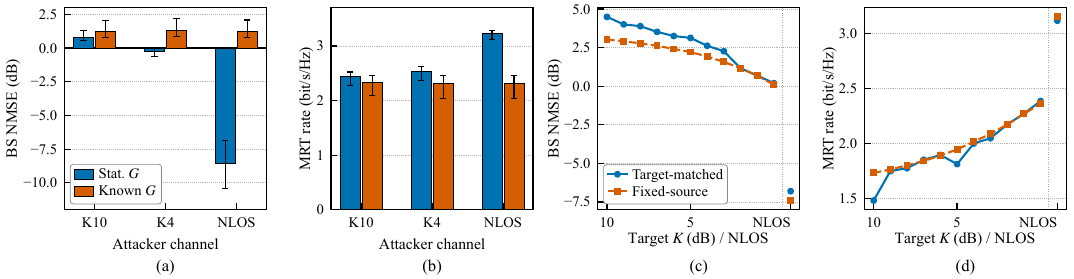}
\caption{Attacker-channel knowledge and statistical transfer. (a), (b) BS NMSE and MRT rate at PSR $-4$~dB with statistical or instantaneous channel knowledge; both use the same ten channel realizations, averaged within each training trial, and bars show means and observed ranges across three trials. (c), (d) BS NMSE and MRT rate for target-matched and fixed-source APGs; the source uses $K=2$~dB and PSR $-2$~dB, with its parameters and transmit PSR held fixed during transfer. Both attacks use PSR $-2$~dB. Curves use one independent training trial. NLOS has no direct component.}
\label{fig:knowledge_transfer}
\end{figure*}

Propagation determines how strongly this threat appears. At PSR 0~dB, the sweep in Fig.~\ref{fig:attackresults} gives APG NMSE values of 7.06, 6.04, and $-6.26$~dB in K10, K4, and NLOS, respectively. The average attacker-link gain is fixed across these distributions. Under~\eqref{eq:rician}, K10 and K4 retain a deterministic spatial component across channel draws, whereas pure NLOS contains scattered components alone. The stronger statistical attack in K10 and K4 is consistent with the generator exploiting this persistent propagation structure. The channel-knowledge comparison below tests how the result changes when the particular propagation realization becomes available.

\subsection{Channel Knowledge and Attack Transfer}
\label{subsec:attack_transfer}

The information comparison isolates the value of knowing the instantaneous attacker channel. In Fig.~\ref{fig:knowledge_transfer}(a) and (b), statistical APG and separately optimized channel-known APGs are evaluated on the same ten attacker-channel realizations in each condition. Results are averaged over this common bank in each trial. The channel-known APG is optimized for its assigned propagation realization, while the current legitimate CSI remains unavailable at deployment. The comparison therefore isolates attacker-channel knowledge while keeping the current legitimate CSI unavailable.

Statistical channel knowledge is sufficient to produce substantial reconstruction errors in K10 and K4. At PSR $-4$~dB, statistical APG gives BS NMSE values of 0.82, $-0.25$, and $-8.59$~dB in K10, K4, and NLOS. Instantaneous channel knowledge increases these to 1.27, 1.35, and 1.27~dB, respectively. The common evaluation bank attributes these differences to the information and corresponding optimization available to the attacker, rather than different propagation samples.

The value of instantaneous information depends on propagation. In NLOS, MRT rates under statistical and channel-known attacks are 3.223 and 2.320~bit/s/Hz, respectively, compared with 2.433 and 2.337~bit/s/Hz in K10. A distribution-based generator can exploit the persistent direct component in K10; knowing a particular NLOS realization additionally exposes the spatial and multipath structure that varies across draws. This interpretation is consistent with the much larger information benefit in NLOS. Statistical knowledge and instantaneous knowledge consequently describe different threats even when their average link gain and transmit-energy budget coincide.

Statistical mismatch poses a separate deployment question. Fig.~\ref{fig:knowledge_transfer}(c) and (d) evaluate an APG trained at $K=2$~dB and PSR $-2$~dB across other attacker channels. Its parameters and transmit PSR remain fixed; target-matched APGs are optimized under each target distribution. The sweeps show that the fixed-source attack retains the main reconstruction and MRT trends as the target $K$ changes. In this model, $K=0$~dB still contains a direct component, while NLOS is the separate condition with no direct component.

Across the repeated trials at representative target conditions, transferred and target-matched APGs give NMSE values of 2.04 and 2.83~dB in K10, and 1.33 and 1.64~dB in K4, respectively. These statistics complement the propagation trends in Fig.~\ref{fig:knowledge_transfer}(c) and (d). Matching the target statistics increases reconstruction damage at these conditions, while the fixed-source generator remains disruptive without retraining. The corresponding K10 MRT rates are 2.078 and 2.006~bit/s/Hz, confirming that both attacks impair directional use of the reconstructed CSI.

In NLOS, the transferred and matched NMSE values are $-8.86$ and $-7.53$~dB, with MRT rates of 3.233 and 3.186~bit/s/Hz. Both retain more useful directional information than under the considered K10 and K4 attacks. Similar transfer and matched performance must therefore be read together with the absolute reconstruction and MRT values. Taken with Fig.~\ref{fig:knowledge_transfer}(a) and (b), the results distinguish two effects: instantaneous channel knowledge can sharply strengthen NLOS attacks, while statistical mismatch leaves a substantial pilot threat in K10 and K4.

\subsection{Mitigation of the Pilot Injection Attack}
\label{subsec:adaptive_defense}

\begin{figure*}[!t]
\centering
\includegraphics[width=\textwidth]{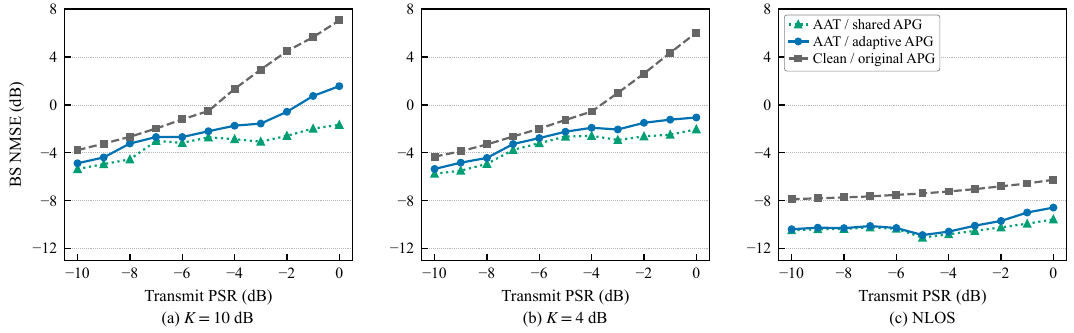}
\caption{BS NMSE under (a) K10, (b) K4, and (c) NLOS attacker channels. Each AAT receiver is trained at its evaluated channel and PSR. The shared APG is optimized against Clean; the adaptive APG is reoptimized against AAT. Curves use one independent training trial; Table~\ref{tab:defense} summarizes three trials.}
\label{fig:defenseresults}
\end{figure*}

Feedback network training is used as a countermeasure to the identified attack path. Condition-matched AAT can improve the final receiver's CSI reconstruction under attack. Fig.~\ref{fig:defenseresults} compares Clean and AAT under shared and adaptive APGs. The shared attack keeps the APG optimized for Clean fixed and isolates the receiver's response to the same disturbances. The adaptive attack reoptimizes APG against each fixed final AAT receiver and tests robustness after attacker adaptation. This distinction connects observed robustness with the attacker that responds to protected mapping.

\begin{table*}[!t]
\caption{CSI Reconstruction and MRT Performance Under Adaptive Pilot Attacks}
\label{tab:defense}
\centering
\small
\setlength{\tabcolsep}{4pt}
\begin{tabular}{@{}llcccrcc@{}}
\toprule
Attacker & PSR & \multicolumn{3}{c}{Linear NMSE (mean $\pm$ standard deviation)} & APG gain & \multicolumn{2}{c}{MRT rate under APG}\\
channel & (dB) & Clean / APG & AAT / adaptive APG & AAT / adaptive UAP-PGD & (dB) & Clean & AAT\\
\midrule
K10 & -4 & $1.214 \pm 0.126$ & \textbf{0.587} $\pm$ \textbf{0.072} & $0.622 \pm 0.073$ & \textbf{3.15} & 2.429 & \textbf{2.586}\\
K10 & 0 & $2.912 \pm 1.880$ & \textbf{1.368} $\pm$ \textbf{0.069} & $0.943 \pm 0.162$ & \textbf{3.28} & 1.684 & \textbf{2.251}\\
K4 & -4 & $0.956 \pm 0.066$ & \textbf{0.534} $\pm$ \textbf{0.095} & $0.529 \pm 0.098$ & \textbf{2.53} & 2.527 & \textbf{2.555}\\
K4 & 0 & $2.428 \pm 1.380$ & \textbf{1.014} $\pm$ \textbf{0.201} & $0.908 \pm 0.123$ & \textbf{3.79} & 1.822 & \textbf{2.294}\\
NLOS & -4 & $0.145 \pm 0.038$ & \textbf{0.083} $\pm$ \textbf{0.003} & $0.081 \pm 0.004$ & \textbf{2.40} & 3.217 & \textbf{3.293}\\
NLOS & 0 & $0.223 \pm 0.014$ & \textbf{0.121} $\pm$ \textbf{0.015} & $0.110 \pm 0.003$ & \textbf{2.65} & 3.144 & \textbf{3.244}\\
\bottomrule
\end{tabular}\space\null
\par\smallskip
\begin{minipage}{.98\textwidth}
\footnotesize The three NMSE columns report linear means and sample standard deviations. APG gain is the reduction in NMSE computed from the ratio of linear means. MRT rates are in bit/s/Hz. Each final receiver needs to be evaluated with an adaptive attacker. The bold entries show the AAT NMSE, gain and MRT rate of paired APG comparisons.
\end{minipage}
\end{table*}

The curves show a benefit from condition-matched AAT across the evaluated PSR range, and Table~\ref{tab:defense} provides the corresponding comparison at six channel--PSR conditions. Against adaptive APG, AAT reduces NMSE by 2.40--3.79~dB relative to Clean across these conditions. At K10 and PSR $-4$~dB, NMSE decreases from 0.84 to $-2.31$~dB; at PSR 0~dB, it decreases from 4.64 to 1.36~dB. K4 and NLOS have improvements, and the defense benefit extends to the considered propagation structure.

Attack adaptation changes the magnitude of the observed gain. At K10 and PSR $-4$~dB, the original APG gives NMSE values of 0.84~dB on Clean and $-3.55$~dB on AAT. Reoptimizing APG against AAT raises its NMSE to $-2.31$~dB. The adaptive attacker is more destructive than the shared one, yet AAT retains a 3.15-dB improvement over Clean. The two evaluations consequently reveal both the response to a fixed attack and the robustness remaining after attack adaptation.

The defended receiver is also evaluated with a different end-to-end attack. At K10 and PSR $-4$~dB, AAT attains an NMSE of $-2.06$~dB under reoptimized UAP-PGD, close to the adaptive-APG result. The APG and UAP-PGD columns in Table~\ref{tab:defense} report attacks optimized independently for the final AAT receiver across all six conditions. These columns assess the defended mapping through different waveform parameterizations; Fig.~\ref{fig:ablation}(b) gives the paired Clean--AAT comparison under UAP-PGD with matched feedback training.

Reconstruction improvements are accompanied by higher mean MRT rates. Under APG, the K10 MRT rate increases from 2.429 to 2.586~bit/s/Hz at PSR $-4$~dB and from 1.684 to 2.251~bit/s/Hz at PSR 0~dB. The mean K4 rate likewise rises from 1.822 to 2.294~bit/s/Hz at PSR 0~dB. In NLOS, the corresponding increase is from 3.144 to 3.244~bit/s/Hz, where Clean already retains a higher rate under the statistical attack. The size of rate recovery relies on the directional damage before defense and the reduction of NMSE.

The following comparison is carried out to see which parts of the adopted training plan are helpful for mitigation under common feedback training.

\subsection{Countermeasure Design and Performance Tradeoffs}
\label{subsec:ablation}

\begin{figure*}[!t]
\centering
\includegraphics[width=\textwidth]{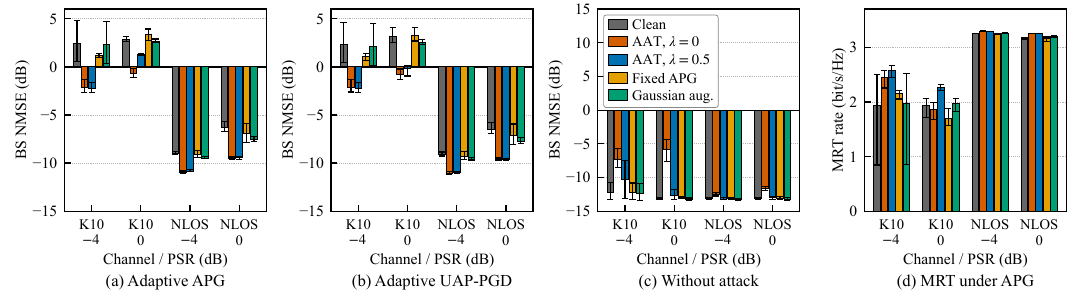}
\caption{Ablation under common feedback-training conditions: (a) adaptive-APG NMSE, (b) adaptive-UAP-PGD NMSE, (c) NMSE without attack, and (d) MRT rate under APG. The horizontal-axis PSR specifies attack strength during training; no injection is applied in (c). Bars and error bars denote the mean and observed range across independent training trials.}
\label{fig:ablation}
\end{figure*}

The training controls examine how updating APG during receiver training and retaining a clean loss affect the defense. Fig.~\ref{fig:ablation} compares Clean, adversarial training with a fixed APG, Gaussian augmentation, and AAT with $\lambda=0$ and $0.5$. Here, $\lambda$ is the clean-loss weight in~\eqref{eq:aat}. Under the common feedback-training protocol in Section~\ref{subsec:simulation_setup}, APG and UAP-PGD are reoptimized against every final receiver. Model selection weights clean and attacked validation losses equally.

The fixed-APG control exposes the receiver to an optimized but stationary attack, while Gaussian augmentation introduces unstructured disturbances. Both use equal clean and attacked loss weights, matching AAT with $\lambda=0.5$ in this respect. Their comparison with AAT tests whether adapting APG to the changing receiver improves the resulting defense. Comparing the two AAT weights then isolates the clean reconstruction term under the same alternating mechanism; $\lambda=0$ uses the attacked loss alone.

Updating APG during feedback-network training yields lower reconstruction errors under adaptive attacks at K10 and PSR $-4$~dB. Clean, fixed-APG training, and Gaussian augmentation give APG NMSE values of 2.46, 1.15, and 2.32~dB, respectively, whereas AAT attains $-2.11$~dB with $\lambda=0$ and $-2.25$~dB with $\lambda=0.5$. The latter improves on Clean and fixed-APG training by 4.72 and 3.40~dB, respectively. Under UAP-PGD, NMSE similarly decreases from 2.35~dB for Clean to $-2.28$~dB for AAT with $\lambda=0.5$. With feedback-network training matched across strategies, the separation from fixed-APG training supports adapting the adversary jointly with the receiver.

Clean-loss mixing improves normal-operation accuracy relative to attacked-only training and regulates its tradeoff with attacked reconstruction and beamforming. At the same K10, $-4$-dB condition, increasing $\lambda$ from 0 to 0.5 improves clean NMSE from $-7.33$ to $-10.29$~dB while changing adaptive-APG NMSE from $-2.11$ to $-2.25$~dB. The Clean reference gives $-12.28$~dB without injection, placing the two weights in the context of normal-operation reconstruction. The clean term recovers 2.96~dB of accuracy relative to attacked-only AAT. Meanwhile, APG SGCS rises from 0.484 to 0.540 and MRT rate from 2.452 to 2.572~bit/s/Hz, so the improvement in clean accuracy is accompanied by improved directional performance under attack at this condition.

At the stronger K10, 0-dB training condition, the two weights yield different reconstruction--beamforming tradeoffs. AAT with $\lambda=0$ gives a lower APG NMSE of $-0.72$~dB than the 1.31~dB obtained with $\lambda=0.5$. The latter, however, increases SGCS from 0.301 to 0.425 and MRT rate from 1.871 to 2.266~bit/s/Hz, while improving clean NMSE from $-5.89$ to $-12.70$~dB; Clean gives $-13.15$~dB without attack. A lower total squared error need not preserve spatial orientation equally well. The mixed objective can improve beamforming performance and close-reference clean reconstruction; the attacked objective favors the NMSE metric. Therefore, the appropriate weight needs to be determined according to the reconstruction and transmission needs.

In NLOS, AAT improves robustness while maintaining clean accuracy. At PSR $-4$~dB, AAT with $\lambda=0.5$ reduces APG NMSE from $-9.00$ to $-10.83$~dB, while clean NMSE changes from $-13.17$ to $-13.18$~dB. At PSR 0~dB, attacked NMSE decreases from $-6.32$ to $-9.44$~dB, with clean NMSE values of $-13.14$ and $-13.03$~dB for Clean and AAT. APG MRT also rises from 3.162 to 3.252~bit/s/Hz at the latter condition. The weaker statistical attack makes more directional information available; the mixed objective retains clean reconstruction with a small change in the rate.

The fixed attack and augmentation control support the adaptation of APG in the adopted training plan. The clean term determines the balance between attacked reconstruction, normal operation accuracy, and beamforming directional information. Combined with the attack-impact and channel-knowledge results, these findings show both the consequences of the pilot-injection path and the feasibility of mitigating it through feedback training. Acquisition integrity matters even when the resulting feedback payload is delivered correctly.

\section{Conclusion}
This paper has identified and evaluated a pilot-injection attack path against AI-based quantized CSI feedback. Contaminated acquisition can yield correctly delivered bits that carry inaccurate channel information. Modeling propagation, LMMSE estimation, compression, and quantized reconstruction connects the injection to BS reconstruction and beamforming. APG and universal-grid optimization implement this path; comparison with estimator-targeted injection demonstrates the role of the final feedback mapping. Statistical injection is relatively disruptive in the K10/K4 conditions without attacker access to current legitimate CSI or instantaneous attacker-channel observations. AAT trains the feedback encoder and decoder on contaminated pilots as a countermeasure. With training matched to each evaluated channel--power condition, it reduces mean NMSE by 2.40--3.79~dB against reoptimized APGs and increases mean MRT rate across six representative conditions. The training controls show how attacker adaptation during training and clean-loss weighting affect the reconstruction--beamforming balance. These findings establish the importance of acquisition integrity alongside payload delivery in securing AI-based CSI feedback.

\bibliographystyle{IEEEtran}
\bibliography{PSAA_reference_canonical}
\end{document}